\documentclass[aps,prl,twocolumn,groupedaddress]{revtex4}
\usepackage{graphicx}
\begin{document}

\title{Geodesic Nucleation and Evolution of a de-Sitter Brane}

\author{Aharon Davidson, David Karasik and Yoav Lederer}
\email[Email:~]{davidson@bgu.ac.il}
\homepage[HomePage:~]{www.bgu.ac.il/~davidson}
\affiliation{Physics Department, Ben-Gurion University, Beer-Sheva
84105, Israel}

\date{March 15, 2005}

\begin{abstract}
	Within the framework of Geodesic Brane Gravity, the deviation
	from General Relativity is parameterized by the conserved bulk
	energy.
	The corresponding geodesic evolution/nucleation of a de-Sitter
	brane is shown to be exclusively driven by a double-well
	Higgs potential, rather than by a plain cosmological constant.
	The (hairy) horizon serves then as the locus of unbroken $Z_{2}$
	symmetry.
	The quartic structure of the scalar potential, singled out on
	finiteness grounds of the total (including the dark component)
	energy density, chooses the Hartle-Hawking no-boundary
	proposal.  
\end{abstract}
 
\pacs{04.50+h, 95.35+d, 11.27+d}
\maketitle

The Randall-Sundrum model\cite{RS} has in fact re-ignited the interest
in brane gravity.
A completely different approach, to be referred to as Geodesic Brane
Gravity (GBG), has been advocated long ago by Regge-Teitelboim\cite{RT}.
Originally aiming towards Quantum Gravity, the RT theory has been
proposed with the motivation that the first principles which govern the
evolution of the entire Universe need not be too different from those
which determine the world-line (world-sheet) behavior of a point particle
(elementary string).
Following RG, the brane Universe can be regarded as a $4$-dim extended
object\cite{E,GW}, parameterized by means of $x^{\mu}\,(\mu=0,1,2,3)$,
geodesically floating in some fixed higher dimensional background
spanned by $y^{A}\,(A=0,1,\ldots,N-1)$.
The idea was criticized\cite{Deser} in the past on gauge dependence
grounds, but has been re-considered (in a flat background) by several authors\cite{RTmodels,AD} since.
Being accustomed to General relativity (GR), one may find the RT action
deceptively conventional
\begin{equation}
	S=\int_{}^{}(-\frac{1}{16\pi G}
	{\mathcal{R}}+{\mathcal{L}}_{m})\sqrt{-g}~d^{4}x ~.
	\label{Lagrangian}
\end{equation}
However, in the spirit of classical particle/string theory, it is now the
embedding vector $y^{A}(x^{\mu})$, rather than the induced metric
tensor $g_{\mu\nu}=\eta_{AB}(y)y^{A}_{,\mu}y^{B}_{,\nu}$, which is
unconventionally elevated to the level of the canonical gravitational
field.
Whereas Einstein equations are supposed to get drastically modified,
it is crucial to emphasize that all other equations of motion (variations
with respect to the matter fields) remain absolutely intact.
The corresponding RT gravitational field equations, a weaker system
(only six independent equations) in comparison with Einstein equations,
take the compact form
\begin{equation}
    E^{\mu\nu}\left(y^{A}_{;\mu\nu}+
    \Gamma^{\,A}_{BC}y^{B}_{,\mu}y^{C}_{,\nu}\right)=0 ~,
\end{equation}
where he Einstein tensor
\begin{equation}
    E^{\mu\nu}\equiv\frac{1}{8\pi G}
    \left(R^{\mu\nu}-\frac{1}{2}g^{\mu\nu}R\right)-T^{\mu\nu}
\end{equation}
keeps track of the underlying Einstein-Hilbert Lagrangian.
Invoking the extrinsic curvatures $K^{i}_{\mu\nu}$, the above equations
can be alternatively written in a more geometrically oriented form, namely
$E^{\mu\nu}K^{i}_{\mu\nu}=0$.
The RT formalism has two important features:
\begin{itemize}
\item Every solution of GR equations, namely $E^{\mu\nu}=0$,
	is necessarily a solution of the corresponding RT equations.
	In other words, RT-gravity exhibits a built in GR-limit.
\item Owing to the powerful identity of geometrical origin
	$\eta_{AB}y^{A}_{,\mu}\left(y^{B}_{;\nu\lambda}+
	\Gamma^{\,B}_{CD}y^{C}_{,\nu}y^{D}_{,\lambda}\right)=0$,
	one still recovers energy-momentum conservation
	$T^{\mu\nu}_{\,;\nu}=0$.
\end{itemize}

\medskip
Within the framework of geodesic brane cosmology,
formulated by virtue of $5$-dimensional local isometric
embedding, only a single independent RT-equation survives,
namely
\begin{equation}
    \frac{d}{dt}\left(\sqrt{-g}E^{tt}\dot{y}^{0}\right)=0 ~.
\end{equation}
A trivial integration gives then rise to
\begin{equation}
    \rho a^{3}(\dot{a}^{2}+k)^{1/2}-
    3a(\dot{a}^{2}+k)^{3/2}=-\frac{\omega}{\sqrt{3}} ~,
\end{equation}
accompanied by $\displaystyle{\dot{\rho}+3\frac{\dot{a}}{a}(\rho+P)=0}$.
The constant of integration $\omega$, recognized as
the conserved bulk energy conjugate to the cyclic embedding time
coordinate $y^{0}(t)$, parameterizes the deviation from the Einstein
limit (where $\dot{a}^{2} + k\rightarrow\frac{1}{3}\rho a^{2}$).
A physicist equipped with the traditional Einstein formalism,
presumably unaware of the underlying RT physics, would naturally
re-organize the latter equation into
\begin{equation}
    \dot{a}^{2}+k =
    \frac{1}{3}\left(\rho+\Delta\rho\right)a^{2} ~,
\end{equation}
squeezing all 'anomalous' pieces into $\Delta\rho$.
Our physicist may rightly conclude\cite{AD} that the FRW
evolution of the Universe is governed by the effective energy density
$\rho+\Delta\rho$ rather than by the primitive $\rho$, and
thus may further identify (or just use the language of)
$\Delta\rho \equiv\rho_{dark}$.
A simple algebra reveals the cubic consistency relation
\begin{equation}
    \left(\rho+\rho_{dark}\right)\rho_{dark}^{2}=
    \frac{\omega^{2}}{a^{8}} ~.
    \label{rhod}
\end{equation}
In this paper, we focus attention on the prototype case involving a
minimally coupled scalar field $\phi(t)$, subject to the standard
equation of motion
\begin{equation}
    \ddot{\phi}+3\frac{\dot{a}}{a}\dot{\phi}+
    \frac{dW(\phi)}{d\phi} = 0 ~,
\end{equation}
involving some (yet unspecified) scalar potential $W(\phi)$.

\medskip
Two exclusive features of Geodesic Brane Cosmology, relevant
for our discussion, are worth noting, namely
\begin{itemize}
\item \textit{Positive definite total energy density:}
	To be specific, Eq.(\ref{rhod}) tells us that $\rho_{total}
	\equiv \rho+\rho_{dark} \geq 0$.
	Curiously, this conclusion holds even for $\rho<0$.
\item \textit{Cosmic duality:}
	FRW cosmological evolution cannot tell the configuration 	
	$\{\rho,\,\rho_{dark}\}$
	from its dual $\{\rho+2\rho_{dark},\,-\rho_{dark}\}$, both
	sharing a common $\rho_{total}$.
	A pedagogical example can be provided by the 'empty'
	$\rho=0$ case, whose dual happens to constitute a scalar
	field theory governed by a quintessence-type potential.
\end{itemize}

\medskip
To uncover the mysteries of the so-called dark component $\rho_{dark}$,
we start by asking a simple minded question:
Under what conditions can we obtain eternal deSitter evolution?
It is well known that Einstein's GR requires the introduction of a
positive cosmological constant $\rho=\Lambda>0$.
Counter intuitively, however, within the framework of RT-gravity, the
presence of a scalar field appears mandatory for deriving the exact
DeSitter solution.
Adopting (say) the spatially-closed $k>0$ case, it remains to find the
tenable scalar potential $W(\phi)$, if any, capable of supporting
$\rho_{total}=\Lambda>0$.

\medskip
Guided by eq.(\ref{rhod}), the thing to notice now is the split
\begin{equation}
	\begin{array}{rcl}
	&  \rho=\displaystyle{\Lambda +
	\frac{\omega}{\Lambda^{1/2}a^{4}}}~, & \vspace{6pt}\\
	& \rho_{dark}=\displaystyle{-\frac{\omega}{\Lambda^{1/2}a^{4}}}~. &
	\end{array}
\end{equation}
Differentiating $\rho= \frac{1}{2}\dot{\phi}^{2} + W(\phi)$, we substitute
$\displaystyle{\frac{dW}{d\phi}+\ddot{\phi}}~$ by
$\displaystyle{-3\frac{\dot{a}}{a}\dot{\phi}}~$, to learn that
\begin{equation}
    \dot{\phi}^{2} = \frac{4\omega}{3\Lambda^{1/2}a^{4}} ~,
\end{equation}
and appreciate the fact that associated with our $\omega>0$ is a
\textit{negative} dark energy component $\rho_{dark}<0$ (we recall
in passing the existence of a yet unspecified $\rho_{dark}>0$ dual
theory with identical brane evolution).
We also find that
\begin{equation}
    W = \Lambda +\frac{\omega}{3\Lambda^{1/2}a^{4}} ~,
\end{equation}
and would like, in search of a differential equation for $W(\phi)$,
to also express $\displaystyle{\frac{dW}{d\phi}}$ as a parametric
function of $a$.
To do so, we calculate $\ddot{\phi}$ and plug the result into
the scalar field equation.
We find
\begin{equation}
    \frac{dW}{d\phi} = \pm \frac{1}{a^{3}}
    \sqrt{\frac{4\omega}{3\Lambda^{1/2}}
    (\frac{1}{3}\Lambda a^{2}-k)} ~.
\end{equation}
The next step is to define a function $f(\phi)$,
\begin{equation}
    	f(\phi)\equiv \frac{1}{3}\Lambda-
	\sqrt{\frac{3k^{2}\Lambda^{1/2}}{\omega}
	\left(W(\phi)-\Lambda\right)}~,
\end{equation}
which turns out to satisfy the differential equation
\begin{equation}
    \left(\frac{df}{d\phi}\right)^{2} =
    \frac{3k^{2}\Lambda^{1/2}}{\omega}f ~.
\end{equation}
The solution of this differential  equation is rather serendipitous:
The unique scalar potential capable of supporting an inflationary
deSitter brane is a double-well Higgs potential, given explicitly by
\begin{equation}
    \fbox{${\displaystyle W(\phi) =
    \Lambda + \frac{3\Lambda^{1/2}k^{2}}{16\omega}
    \left(\phi^{2} -\frac{4\omega\Lambda^{1/2}}{9k^{2}}
    \right)^{2}}$}
    \label{Higgs}
\end{equation}
Associated with this double-well potential, but relying on certain
creation initial conditions (to be specified soon) is the full $k>0$
solution
\begin{eqnarray}
	a(t) & = & \sqrt{\frac{3k}{\Lambda}}
	\cosh{\sqrt{\frac{\Lambda}{3}}t} ~, 
	\label{solution1}\\
	\phi(t) & = & \sqrt{\frac{4\omega\Lambda^{1/2}}{9k^{2}}}
	\tanh{\sqrt{\frac{\Lambda}{3}}t} ~.
	 \label{solution2}
\end{eqnarray}

On symmetry (and forth coming Euclidean) grounds, we find
it rewarding to follow Hartle and Hawking\cite{HH} and define
the proper scalar field $b(t) \propto a(t)\phi(t)$.
In this language, the cosmological evolution of the system is
described by the hyperbola
\begin{equation}
    a(t)^{2}-b(t)^{2}=\frac{3k}{\Lambda} ~.
    \label{hyperbola}
\end{equation}

The emerging deSitter inflationary scheme, accompanied by
the auxiliary scalar field, deviates conceptually from the
conventional GR prescription.
Created with a finite radius of
$\displaystyle{a_{0}=\sqrt{\frac{3k}{\Lambda}}}$, while sitting at
the top of the potential hill $\displaystyle{W_{0} = \Lambda\left(
1+\frac{\omega\Lambda^{1/2}}{27k^{2}}\right)}$, the exponentially
growing brane slides down the hill towards the absolute minimum
of the theory.
The latter is conveniently located at the Einstein limit
$W_{\infty}=\Lambda$.
The scalar field, at the meantime, recovering from the
non-conventional creation initial conditions
\begin{equation}
    \phi_{0} = 0 ~,\quad
    \dot{\phi}_{0} = \sqrt{\frac{4\omega
    \Lambda^{3/2}}{9k^{2}}} ~,
\end{equation}
grows monotonically on the way to eventually picking up its
vacuum expectation value
\begin{equation}
    \langle\phi\rangle =
    \sqrt{\frac{4\omega\Lambda^{1/2}}{9k^{2}}} \equiv V~.
\end{equation}
Altogether, accompanied by a remarkable seesaw-type
$\rho\leftrightarrow \rho_{dark}$ interplay, deSitter inflation is
described, within the framework of geodesic brane cosmology,
as a spontaneous symmetry breaking process, with GR eventually
recovered at the absolute minimum.
On the practical side, there is no need to artificially engineer the
shape of a slow-rolling scalar potential in order to maximize the
inflation period; in RT-gravity, an ordinary Higgs potential can
produce eternal inflation.

\medskip
Two important remarks are in order:

\noindent (i) For $k<0$, the situation is very much alike.
Truly, this time one faces
\begin{equation}
   	a(t)\sim\sinh{\sqrt{\frac{\Lambda}{3}}t} ~,~~
	\phi(t) \sim \coth{\sqrt{\frac{\Lambda}{3}}t} ~,
\end{equation}
but the Higgs potential stays invariant under $k\rightarrow-k$.
Nucleated with size zero, accompanied by a monotonically
decreasing scalar field, our exponentially growing open
brane slides again towards the $W_{\infty}=\Lambda$
Einstein limit.
However, contrary to the closed $k>0$ case where only the
\textit{inner} section $(0\leq\phi\leq V)$ of the potential was
involved, it is the \textit{outer} section $(V\leq\phi<\infty)$
which participates in the $k<0$ game.
For $k=0$, the situation is less complicated, with the Higgs
potential reducing to a simple mass term.

\medskip
\noindent (ii) The deSitter metric can also take the static radially
symmetric form
\begin{equation}
    ds^{2} = -\left(1-\textstyle{\frac{1}{3}}
    \Lambda R^{2}\right)dT^{2} +
    \frac{dR^{2}}{\left(1-\frac{1}{3}\Lambda R^{2}\right)} +
    R^{2}d\Omega^{2} ~,
\end{equation}
exhibiting an event horizon at
$\displaystyle{R=\sqrt{\frac{3}{\Lambda}}}$.
Reflecting the $\rho\leftrightarrow \rho_{dark}$ seesaw interplay
between the primitive and the dark energy densities, the auxiliary
scalar field plays in this coordinate system an apparently paradoxical
non-static role.
To see the point, consider (say) the patch
$\displaystyle{R\leq\sqrt{\frac{3}{\Lambda}}}$ covered by
\begin{mathletters}
\begin{eqnarray}
    & \displaystyle{\sqrt{\frac{\Lambda}{3k}}R =
        r \cosh\sqrt{\frac{\Lambda}{3}}t ~,}& \\
    & \displaystyle{\coth \sqrt{\frac{\Lambda}{3}}T =
    \sqrt{1-kr^{2}} \coth\sqrt{\frac{\Lambda}{3}}t ~.}&
\end{eqnarray}
\end{mathletters}
In this coordinate system, the auxiliary $T$-dependent scalar
field acquires the form
\begin{equation}
    \fbox{$\displaystyle{\phi(T,R) = \frac
    {V\sqrt{1-\frac{1}{3}\Lambda R^{2}}
    \sinh\sqrt{\frac{\Lambda}{3}}T}
    {\sqrt{1+\left(1-\frac{1}{3}\Lambda R^{2}\right)
    \sinh^{2}\sqrt{\frac{\Lambda}{3}}T}}}$}
\end{equation}
giving rise to double-kink configuration (a kink-antikink
configuration for $\displaystyle{R\geq\sqrt{\frac{3}{\Lambda}}}$)
scalar hair.
In almost every point $R$ in space, elegantly avoiding the no-hair 
theorems of GR, the scalar field connects
$\phi(-\infty,R)\rightarrow -V$ with $\phi(\infty,R)\rightarrow V$.
It is exclusively on the event horizon, where the scalar field,
experiencing an infinite gravitational red-shift, gets frozen in its
unbroken phase!
In other words, the hairy event horizon appears as the
locus of unbroken ${\cal Z}_{2}$ symmetry.
The generality of this statement may (or maynot) extend beyond the
scope of the present work.

\medskip
To enter the Euclidean regime we perform the Wick rotation
$\displaystyle{t \rightarrow 
-i\left(\tau-\frac{\pi}{2}\sqrt{\frac{3}{\Lambda}}\right)}$.
The exact $k>0$ solution Eqs.(\ref{solution1},\ref{solution2})
transforms then into
\begin{eqnarray}
	a(t) & \rightarrow & a_{E}(\tau) =
	\sqrt{\frac{3k}{\Lambda}}
	\sin{\sqrt{\frac{\Lambda}{3}}\tau} ~, 
	\label{Esolution1}\\
	\phi(t) & \rightarrow & i\phi_{E}(\tau) =
	i\sqrt{\frac{4\omega\Lambda^{1/2}}{9k^{2}}}
	\cot{\sqrt{\frac{\Lambda}{3}}\tau} ~.
	\label{Esolution2}
\end{eqnarray}
The fact that the scalar field turns purely imaginary puts
us in a less familiar territory, in some sense reminding us of the
Coleman-Lee\cite{CL} scheme.
The imaginary time evolution is then best described by the
circle
\begin{equation}
    a_{E}^{2}(\tau) + b_{E}^{2}(\tau) = \frac{3k}{\Lambda} ~,
    \label{circle}
\end{equation}
recognized as the analytic continuation of eq.(\ref{hyperbola}).
This makes the familiar deSitter Euclidean time periodicity
$\displaystyle{\Delta \tau = 2\pi\sqrt{\frac{3}{\Lambda}}}$ manifest,
and opens the door for a generalized Hawking-Hartle no-boundary
proposal.

\begin{figure}[tbp]
    \begin{center}
	\includegraphics[scale=0.5]{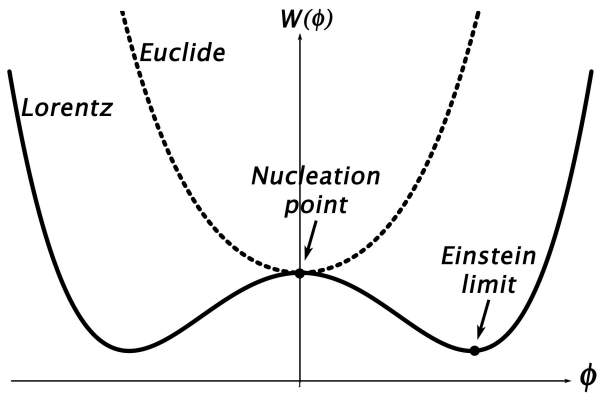}
    \end{center}
    \caption{Geodesic deSitter brane nucleation:
    Evolving from Euclidean no-boundary initial conditions,
    the Euclidean to Lorentzian transition can only occur
    at $\phi=\phi_{E}=0$. GR is asymptotically recovered at
    the Einstein limit of the Lorentzian regime.}
    \label{Fig1}
\end{figure}

\medskip
It is now important to find out which potential actually governs
the imaginary time evolution of $\phi_{E}$?
Traditionally, we have been accustomed to the upside-down
potential $W_{E}(\phi_{E})=-W(\phi_{E})$, but this is definitely
not the case here.
Euclidizing the time derivatives in the scalar field equation,
and simultaneously taking care of $\phi\rightarrow i\phi_{E}$,
brings us back to the original form
\begin{equation}
    \phi_{E}^{\prime\prime}+
    3\frac{a_{E}\prime}{a_{E}}\phi_{E}^{\prime}+
    \frac{\partial W_{E}}{\partial\phi_{E}} = 0 ~,
    \label{Escalar}
\end{equation}
only with $W_{E}(\phi_{E})=+W(i\phi_{E})$.
The resulting potential in the Euclidean regime is then
\begin{equation}
    \fbox{${\displaystyle W_{E}(\phi_{E}) = \Lambda +
    \frac{3\Lambda^{1/2}k^{2}}{16\omega}
    \left(\phi_{E}^{2} +\frac{4\omega\Lambda^{1/2}}
    {9k^{2}}\right)^{2}}$}
\end{equation}
although quartic, this potential is strikingly \textit{not} of
the double-well type.
Furthermore, as depicted in Fig.\ref{Fig1}, the absolute
minimum of $W_{E}(\phi_{E})$ is \textit{tangent} to the
local maximum of $W(\phi)$.
This is by no means coincidental. 
$\phi=\phi_{E}=0$ is the only point where the
Euclidean to Lorentzian transition, to be referred to as brane
nucleation\cite{HLV}, can actually occur.

\medskip
We now attempt to go one step beyond de-Sitter inflation.
To do so, we would like to commit ourselves to a certain
type of scalar potentials, but soon realize that so far
we have not really decoded the principles underlying the
tenable potential $W(\phi)$.
Two related questions are then in order:
\begin{enumerate}
\item Why must $W(\phi)$ exhibit a quartic behavior?
\item Is the quartic potential a mandatory ingredient of geodesic
	 brane cosmology?
\end{enumerate}
The answers to these questions is rooted, quite unexpectedly,
within the Hartle-Hawking no-boundary ansatz\cite{HH}.
We are about to prove, by exclusively predicting a finite
non-vanishing total energy density at the origin, that the quartic 
structure of the potential actually chooses the no-boundary
initial conditions.

\medskip
The smoothness of the Euclidean manifold at the origin dictates
the specific $\tau\rightarrow 0$ behavior $a_{E}(\tau) \simeq
\sqrt{k}\tau$, but may in principle allow for
$\displaystyle{b_{E}(\tau) \simeq \frac{p\sqrt{k}}{\tau^{j-1}}}$.
Now, assuming the asymptotic power behavior 
\begin{equation}
    W(\phi) \simeq \lambda\phi^{N} \quad (\lambda>0) ~,
\end{equation}
the scalar equation of motion eq.(\ref{Escalar}) can
be fulfilled (to the leading order) only provided
\begin{eqnarray}
	& j N=2(j+1) ~,& \\
	& N\lambda p^{(N-2)}+j(j-2)=0 ~.&
\end{eqnarray}
This in turn implies $\rho \sim \tau^{-2(j+1)}$ but
$\rho_{total}\sim\tau^{4(j-1)}$.
Consequently, fully consistent with our expectations, $N$ gets
uniquely fixed by insisting on a finite non-vanishing total energy
density as $a_{E}\rightarrow 0$.
This singles out
\begin{equation}
    	j=1\quad \Rightarrow \quad N=4 ~.
\end{equation}
The special no-boundary initial conditions then read
\begin{equation}
    a_{E} \simeq \sqrt{k}\tau ~, \quad
    b_{E} \simeq \frac{k}{4\lambda} ~,
    \label{initial}
\end{equation}
accompanied by the finite total energy density
\begin{equation}
    \fbox{$\displaystyle{\rho_{total}\simeq
    \left(\frac{4\omega \lambda}{3k^{2}}\right)^{2}}$}
\end{equation}
While the no-boundary initial conditions are
$\omega$-independent, it is the RT bulk energy $\omega$
(used to parameterize the deviation from GR) that actually fixes
the finite value of the total energy density.

\medskip
It is interesting to note that had we carried out a similar
calculation for an $n$-dim brane (we skip the details of the proof),
we would have encountered the famous scale invariant behavior
\begin{equation}
   W(\phi) \sim \phi^{\frac{2}{n}(n+2)}  ~, 
\end{equation}
which happens to be quartic for the special $n=4$ case of interest.
This indicates that, within the framework of geodesic brane cosmology,
there exists a linkage between the apparently disconnected ideas of
Hawking-Hartle no-boundary proposal and global conformal invariance,
pointing presumably towards geodesic dilaton cosmology.

\medskip
Finally, on semi-realistic grounds, while adopting the quartic Higgs
potential, it makes sense to exercise the option of setting
\begin{equation}
	W_{min}(\phi)=0 ~.
\end{equation}
The price for eliminating the residual cosmological constant from
the Einstein limit is a \textit{finite} (yet enhanced in comparison
with standard cosmology) amount of inflation.
On the other hand, subject to the consistent no-boundary initial
conditions eq.(\ref{initial}), we know that the classical Euclidean
evolution is fully determined once the conserved bulk energy
$\omega$ gets specified.
\begin{figure}[tbp]
    \begin{center}
	\begin{tabular}{cc}
	    \includegraphics[scale=0.25]{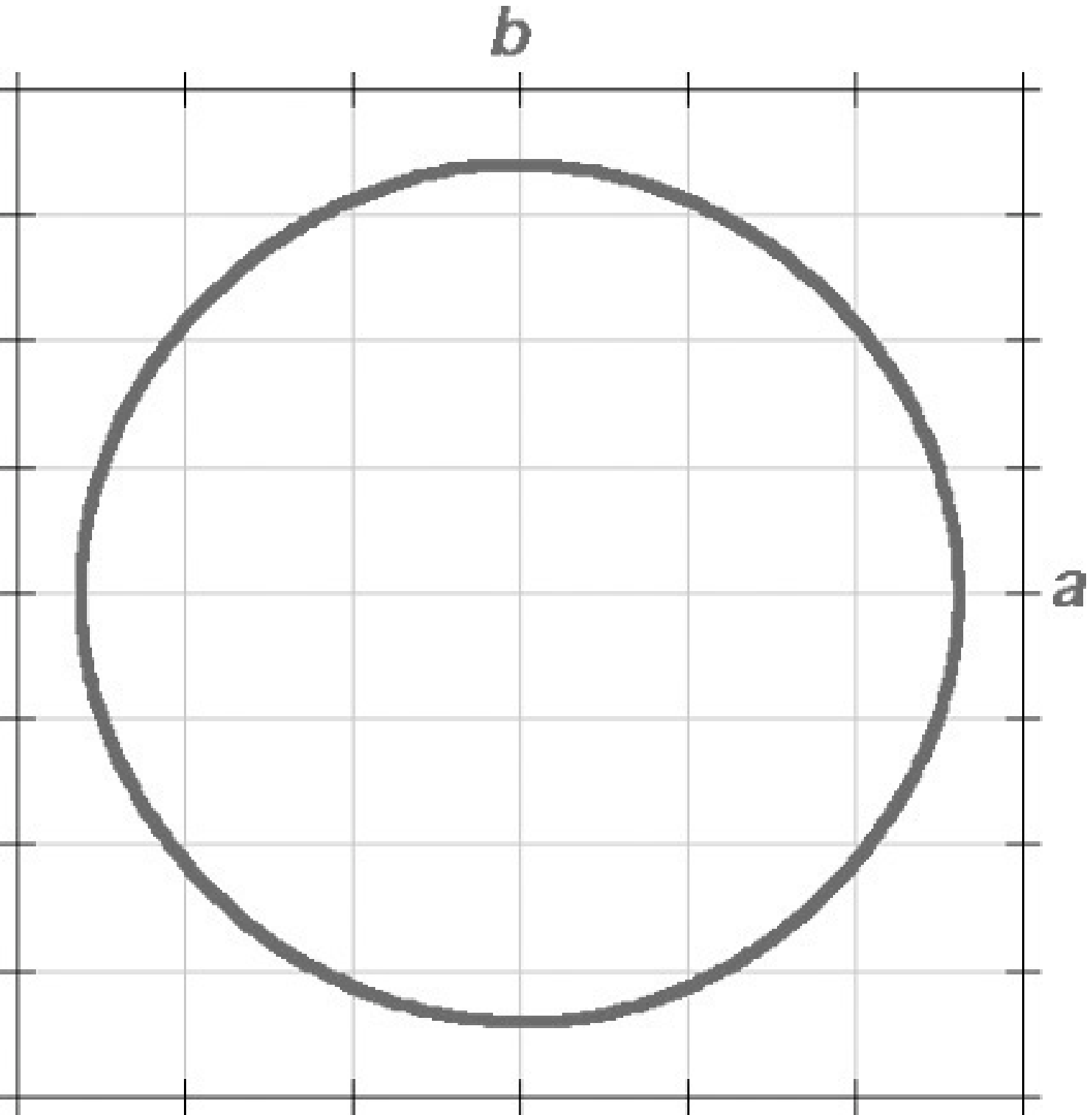} &
	    \includegraphics[scale=0.25]{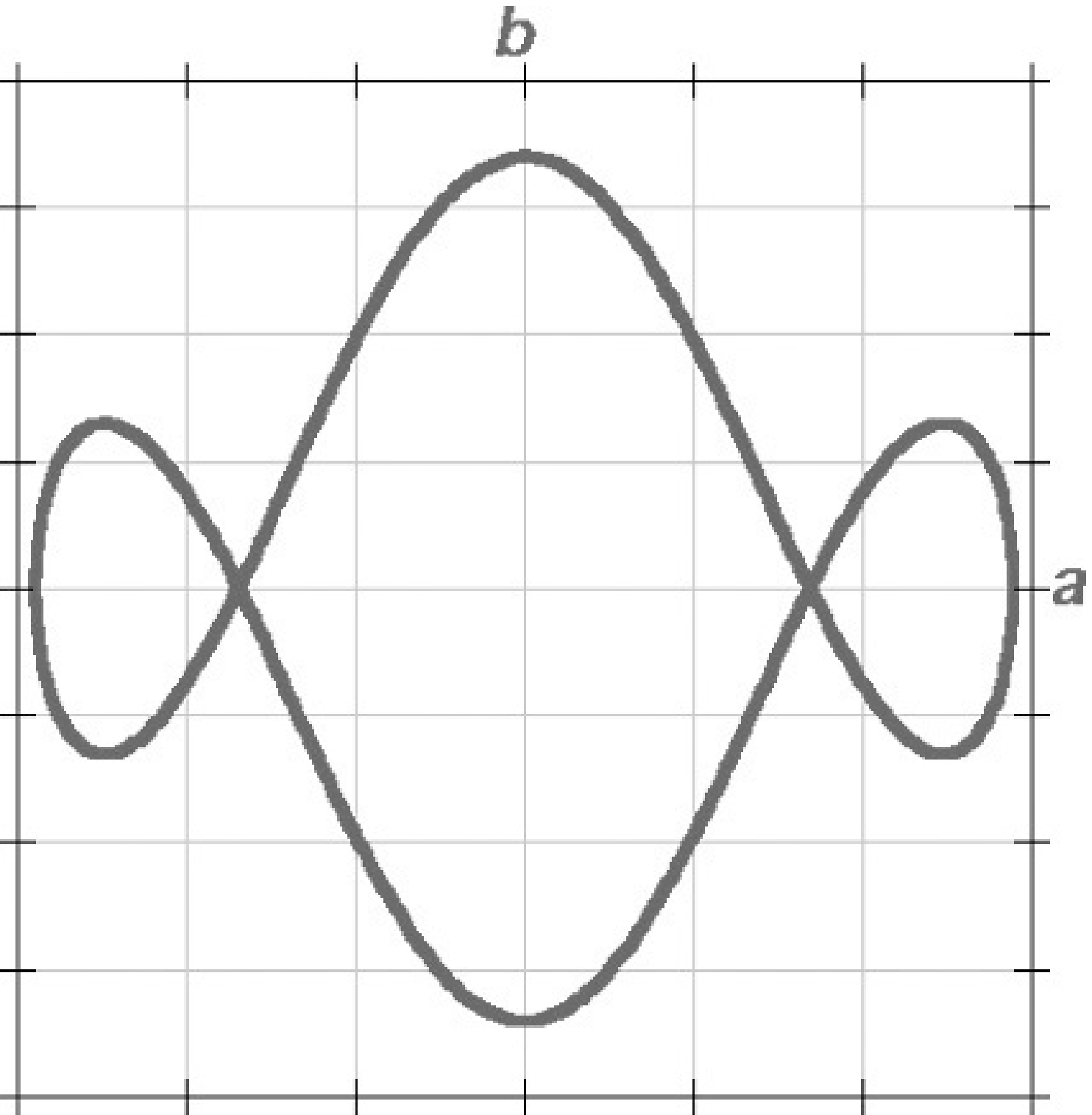}  \\
	    \includegraphics[scale=0.25]{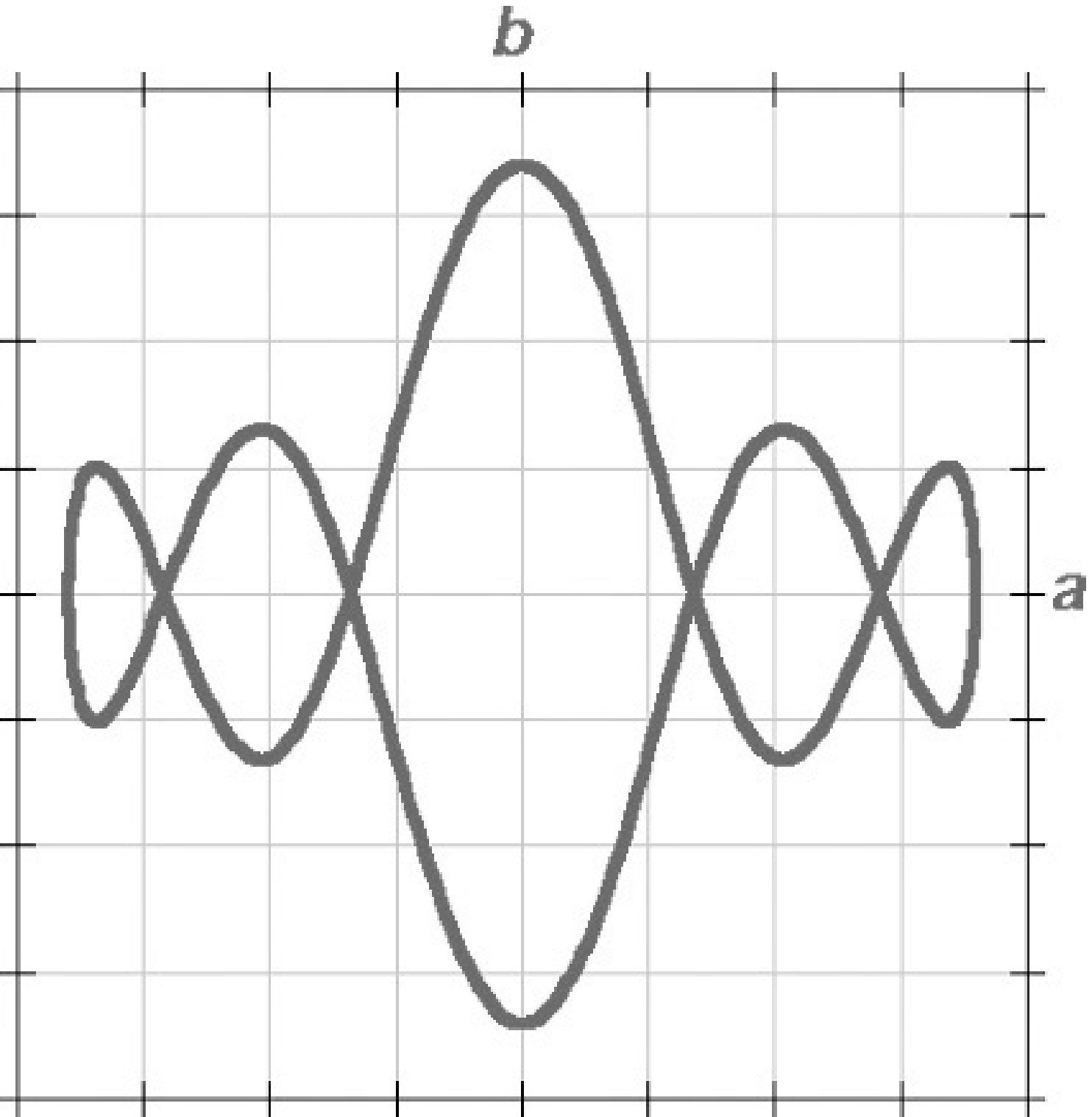} &
	    \includegraphics[scale=0.28]{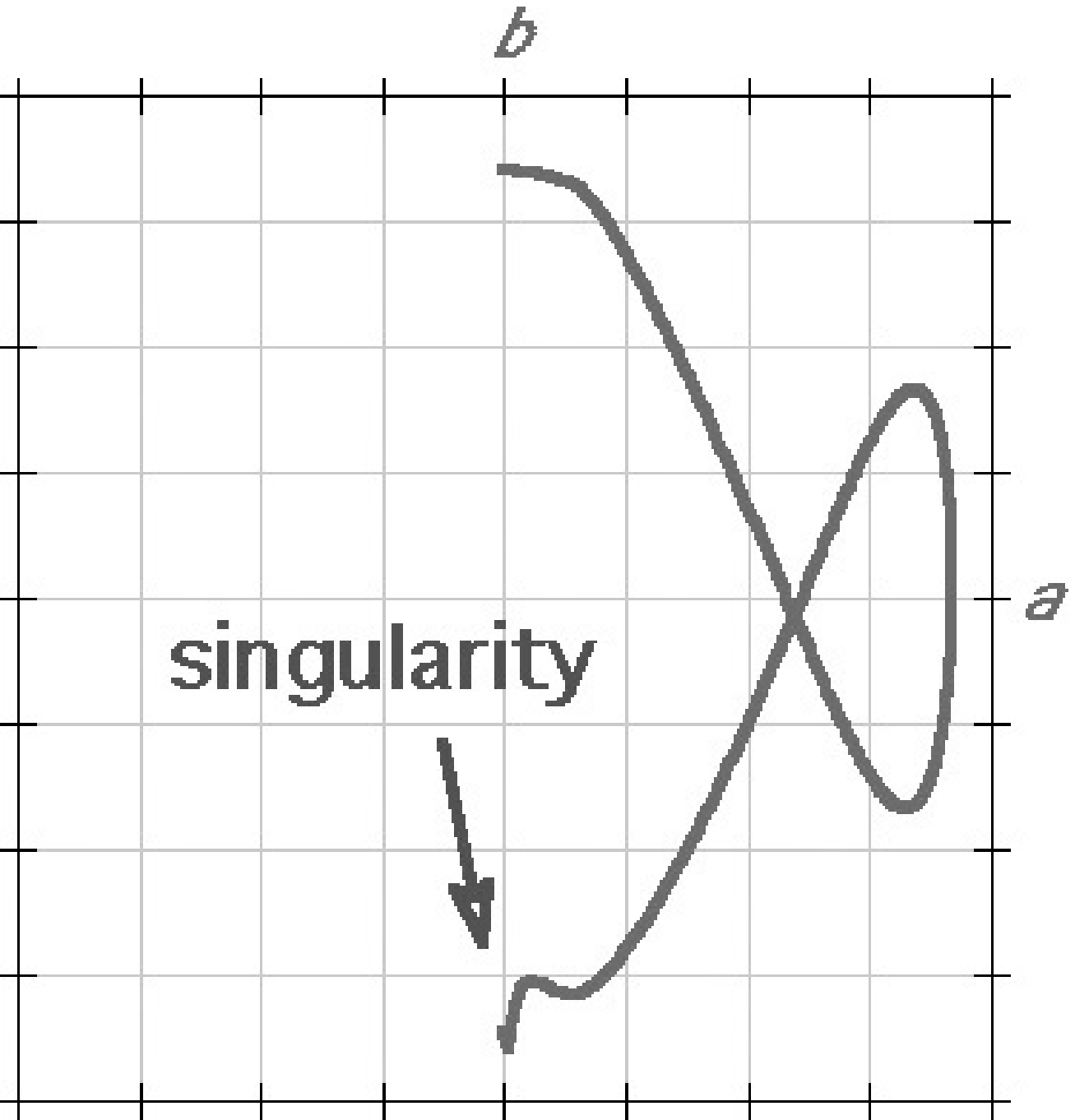}  \\
	\end{tabular}
    \end{center}
    \caption{Imaginary time periodicity and energy density
    regularity are demonstrated, for $\omega=\omega_{1,3,5}$,
    by means of closed Lissajous-like trajectories in the
    $\{a_{E},b_{E}\}$ plane.
    The $\omega=\omega_{2,4,...}$ case is associate with
    Coleman-Dellucia\cite{CD} Eulidization.
    If $\omega\neq\omega_{n}$, a $\cos(\ln\epsilon)$-type
    singularity is developed upon returning to the origin.}
    \label{Fig2}
\end{figure}
Naturally, this provokes a new set of questions:
\begin{enumerate}
\item Does the global structure of the Euclidean manifold still
	exhibit, in some cases, imaginary time periodicity?
\item Under what circumstances, if any, does the total energy
	density evolve free of singularity? If a singularity does occur,
	what is its nature?
\item Can our nucleation conditions $a_{E}^{\prime}=\phi_{E}=0$,
	or else Coleman-Dellucia\cite{CD} conditions
	$a_{E}^{\prime}=\phi_{E}^{\prime}=0$, be met at some finite
	Euclidean time $\tau_{0}$?
\end{enumerate}

\medskip
We claim, skipping the analytic proof (to be published elsewhere)
but based on detailed numerical calculations, that $\tau$-periodicity,
$a_{E},b_{E}$-regularity, and the occurrence of spontaneous
nucleation, share in fact the one and the same origin.
They can all be simultaneously achieved provided
\begin{equation}
	\omega=\omega_{n}
\end{equation}
is properly quantized, in agreement with some previous WKB 
approximation\cite{AD}.
The integer $n$ counts the total number of times the proper
scalar field $b_{E}$ crosses the absolute minimum of the potential
during half a period.
For the $n$-odd case of interest ($n$-even is associated with
Coleman-Dellucia), reflecting the interplay of two periodicities, we
encounter (see Fig.\ref{Fig2}) $n$-loop \textit{closed} trajectories in the
$\{a_{E},b_{E}\}$ plane which resemble the Lissajous figures.
Notice that the Euclidean de-Sitter configuration Eq.(\ref{circle})
clearly belongs to the $n=1$ category.

\end{document}